\documentclass[runningheads]{llncs}
\usepackage[T1]{fontenc}
\usepackage{graphicx}

\usepackage{xspace}
\def\Mathlib{Mathlib\xspace}
\def\mathlib{Mathlib\xspace}

\usepackage[frozencache=true]{minted}
\newcommand{\lean}[1]{\mintinline{lean}{#1}}
\newcommand{\hide}[1]{}

\usepackage{hyperref}
\hypersetup{
    bookmarks,bookmarksopen,bookmarksnumbered,colorlinks,breaklinks,
    colorlinks=true,       % false: boxed links; true: colored links
    linkcolor=green,          % color of internal links
    citecolor=magenta,        % color of links to bibliography
    filecolor=green,      % color of file links
    urlcolor=blue           % color of external links
}
\usepackage[backend=biber,isbn=false,eprint=false,maxbibnames=5,maxalphanames=4]{biblatex}
\newcommand{\ddt}[1]{\marginpar{{\tiny{DT: #1}}}}
\usepackage{color}

\begin{document}
\title{Growing Mathlib: maintenance of a large scale mathematical library}
\author{Anne Baanen\inst{1}\orcidID{0000-0001-8497-3683}\and
Matthew Robert Ballard\inst{2}\orcidID{0000-0001-5819-0159}\and
Johan Commelin\inst{1,3}\orcidID{0009-0000-2025-9771}\and
Bryan Gin--ge Chen\orcidID{0000-0002-7023-7252}\and
Michael Rothgang\inst{4}\orcidID{0009-0001-8307-4122}\and
Damiano Testa\inst{5}\orcidID{0009-0004-5949-6861}}
\authorrunning{A. Baanen et al.}

\institute{Lean FRO, USA \and
University of South Carolina, USA \and
Mathematical Institute, Utrecht University, The Netherlands \and
Mathematical Institute, University of Bonn, Germany \and
Mathematical Institute, University of Warwick, UK}

\maketitle

\begin{abstract}
The Lean mathematical library \mathlib is one of the fastest-growing libraries of formalised mathematics.
We describe various strategies to manage this growth, while allowing for change and avoiding maintainer overload.
This includes dealing with breaking changes via a deprecation system,
using code quality analysis tools (\emph{linters}) to provide direct user feedback about common pitfalls,
speeding up compilation times through conscious library (re-)design,
dealing with technical debt
as well as writing custom tooling to help with the review and triage of new contributions.

\end{abstract}

\section{Introduction}

The mathematical library \Mathlib~\cite{Mathlib}, written in the Lean theorem prover, is perhaps the fastest-growing library of formalised mathematics. Created in 2017, it now contains 1.9 million lines of code across a broad range of subjects.
\Mathlib's focus on being an integrated library for mathematics makes it well-suited for formalising current research mathematics.
Recent formalised results based on \mathlib include a fundamental result of Scholze and Clausen about condensed mathematics~\cite{LTEChallenge,LTEFinal,CommelinTopazAbstractionBoundaries}, Smale's sphere eversion via an $h$-principle~\cite{SphereEversionCPP},
Campos--Griffiths--Morris--Sahasrabudhe's breakthrough on upper bounds for diagonal Ramsey numbers~\cite{MehtaDiagonalRamsey} and
Gowers--Green--Manners--Tao's breakthrough on Marton's polynomial Freiman--Rusza conjecture \cites{PFR1,PFR2,PFRImproved}.
\Mathlib is also crucial for Buzzard's ongoing project \cite{BuzzardFLT} to formalise a proof of Fermat's Last Theorem.
In a different direction, \mathlib has been used in industry research, such as for verifying the correctness of the SampCert differential privacy framework~\cite{SampCert}.

\Mathlib is run by a community of users as an open source project, enabling and encouraging contributions by users with a wide range of backgrounds and expertise. This allows for a broad range of contributions (of formalisations, but also with the infrastructure and other supporting work).
At the same time, this presents challenges around knowledge transfer and ensuring the library stays coherent as a whole.
To address this, \mathlib makes extensive user of \emph{linters}, to help users with intricacies of the Lean proof assistant and enforce certain global coherence properties. (We describe this in Section~\ref{sec:linters}.)

\Mathlib's continued growth also presents other challenges for scaling. Some of these are technical---such as adapting existing code to other changes in the library (Section~\ref{sec:deprecations}), ensuring that \mathlib continues to compile reasonably quickly (see Section~\ref{sec:speed}) and managing technical debt (see Section~\ref{sec:techdebt}).
With any large group, social processes become pertinent (such as, ensuring all communication stays friendly and welcoming (\mathlib has a code of conduct with a dedicated team enforcing it), or keeping some alignment on an overall vision of \mathlib).
A key process in both maintaining a coherent library and teaching new contributors about formalisation best practices is mandatory code review. This process needs to be supported and adapted to scale to the influx of contributions (see Section~\ref{sec:reviewTooling}).

We describe the \mathlib community's strategies for dealing with this growth, focusing on the above aspects in turn.
The use of linters (in a less extensive way) was already described in previous work \cite{VanDoornEbnerLewis20}; the other sections are new.
While some details are specific to Lean and \mathlib, we believe that many of them apply broadly to libraries of formal mathematics knowledge.

\section{Related work}

Various interactive theorem provers have come to enjoy large libraries of research mathematics---each with its own strengths and weaknesses.
The Mizar Mathematical Library \cite{RoleOfMML} is the oldest and perhaps the largest library (at 3.7 millions lines of code (MLOC)) of formalised mathematics.
The Isabelle/HOL system has very powerful automation; the \href{https://www.isa-afp.org/}{Archive of Formal Proofs} (AFP; 4.8 MLOC%
\footnote{The AFP has a broader focus \cite{MiningAFP}, with about half its current articles covering the analysis and verification of particular algorithms. Arguably, this makes the MML a larger library of formalised \emph{mathematics}.}) still has the strongest complex analysis library.
The Rocq theorem prover allows expressing complex mathematical objects (such as schemes, the tangent space of a manifold or vector bundles) more naturally, thanks to its foundations including dependent types.
While its mathematical components library \cite{mathcomp} is significantly smaller (150 000 lines of code), Rocq has been used to formalise major results such as the Four Colour and the Feit--Thompson theorems (which are each of similar size as the \emph{mathematical components}).
Let us also mention the Flyspeck project \cite{FlyspeckComplete} as an example of combining large libraries in two different proof assistants.

The MML introduced mandatory review of contributions in 2006, as well as tooling to check for duplicate lemmas. (Such tooling also exists for Rocq \cite{RocqGoalClone} and Lean \cite{tryAtEachStep}.)
Klein et al have written \cite{KleinLargeScale} about maintaining large-scale Isabelle libraries. They name many challenges similar to ours, including coordination and collaboration in large and diverse projects.
The Isabelle system also employs custom linters \cite{IsabelleLinters} --- which are not mandated for existing entries, but enforced for new submissions.
Zimmermann \cite{ZimmermannThesis} has written about recent changes to Rocq's development processes (including switching to GitHub and a pull-request based model, dealing with breaking changes and the challenges with distributing reviews). There is also custom tooling to support Rocq maintainers \cite{RocqBot}.
Huch's in-progress PhD thesis \cite{HuchThesisDraft} describes efforts to address scaling issues in the AFP, including a discussion of maintainenance efforts.

Growing a large formal library can require changes to the underlying prover.
Such work on Lean is out of scope of this paper; Wenzel discussed it for Isabelle \cite{Wenzel19}.
They also describe tooling to automatically test the impact of core system or theory changes on dependent projects.%
\footnote{Similar tooling for Lean does not exist yet, but is being planned by the Lean FRO.}
Test-case minimisation \cite{RocqMinimiser} can also be instrumental for quickly diagnosing and fixing broken proofs ``downstream''.
Huch and Wenzel have also written about the challenges of maintaining and scaling Isabelle's supporting infrastructure \cite{HuchIsabelleBuilds,HuchWenzelParallelAFP}.

There is also considerable work on proof repair, such as adjusting code to breaking changes.
Luan et al \cite{LuanCompatAFP} have studied and classified compatibility issues in the AFP.
Tan et al \cite{BurdenOfProof} have developed tooling for automatically fixing proofs in Isabelle/HOL.
Gopinathan et al \cite{GopinathanProofRepair} have developed proof repair tooling for the Rocq prover.

Despite the growth of large libraries of formal proofs, there is still relatively little work describing tools and best practices for their maintenance.
Ringer et al \cite{Ringer19} note the gap between proof engineering and software engineering in this respect.

\section{The deprecation system}
\label{sec:deprecations}

Maintaining \mathlib involves changing existing code. Generalizations, refactors, reorganization of content, syntax changes and so on are \emph{exceedingly} common in \mathlib.
As a result, code written for one version of \mathlib may no longer compile with a newer version.
This applies both within \mathlib (contributions to one area need to deal with effects on all other areas), to contributions under review and particularly to projects depending on \mathlib. Many large formalisation projects are carried out in a repository depending on \mathlib. If keeping up with the latest \mathlib changes is too onerous, contributing results back to \mathlib may become impossible.

The spectrum of possible breaking changes is rather diverse, and ranges from one-off changes when upgrading to the next Lean version to adjusting to particular refactorings in \mathlib.
In principle, Lean's excellent meta-programming facilities allow writing custom migrations for virtually all changes, transforming working code from one version to the next.
Theoretically, such migrations could be arbitrarily complicated.
In practice, however, the vast majority of adjustments are very simple to deal with: renaming (the declaration which states) a theorem or definition, renaming files (which requires adjusting \lean{import} statements) and superficial syntax changes. The third item is addressed through dedicated syntax linters (see Section~\ref{sec:linters}); these can also help with semantic changes.

Lean has a deprecation system specifically designed to handle renamings gracefully.
When a declaration is renamed, for a grace period of several months, we add a deprecation marker documenting the rename. This connects the old and new names (and allows providing a custom message to guide the migration process). Upon using a deprecated declaration, the user sees a warning, indicating which declaration to use instead.
This allows for a much smoother migration. % The replacement itself is rather mechanical; there are several prototype scripts which automatically apply them.

Dealing with renamed files is harder.
Lean looks at file imports before the deprecation system (which is implemented as a linter) has the chance to run---so a file that imported the file with the old name would stop building entirely.
The solution is similar: the module with the old name is not erased, but is preserved and stained.
Thus, it can be imported without breaking the build---and a linter suggests to replace the old module name, with appropriate imports.
These could be either the direct imports of that module, in the case of a reorganization of content; or the renamed module, in the case of a simple move.
This informs users of where the declarations that were present in the old file may have ended up.
% Creating module deprecations can likewise be automated.

\section{Semantic linters}
\label{sec:linters}

In general terms, a {\emph{linter}} is a program whose purpose is to ensure that the code in a project satisfies certain requirements.
The term requirements should be interpreted very broadly: a linter could enforce
\begin{itemize}
	\item code style conventions (such as the default indentation of code);
	\item file naming conventions (such as never using paths that differ only in capitalization);
	\item function naming conventions (such as making sure that namespaces in declarations are not repeated);
	\item deprecating commands that have been replaced by newer code;
	% \item restricting axioms (such as enforcing that some declaration does not use the axiom of choice);
\end{itemize}
and in fact \mathlib uses linters for all of these purposes.

To be precise, \Mathlib uses two different linter frameworks, with distinct strengths and weaknesses.
Lean itself provides a \lean{Lean.Elab.Command.Linter} type, which contains the linter's name and a monadic function that takes the \href{https://lean-lang.org/doc/reference/latest//Notations-and-Macros/Defining-New-Syntax}{syntax} of the command as input and can run arbitrary code.\footnote{Technically, a linter function returns a \lean{Unit}, i.e.\ the only interesting outcomes of the linter are its side effects.}
Lean's community-maintained extended library \href{https://github.com/leanprover-community/batteries}{Batteries} provides the parallel structure\\\lean{Batteries.Tactic.Lint.Linter} that is similar to the Lean-native linters, but takes a declaration name as its input and runs code in a different monad.
For the purpose of the present discussion, the main difference between the two kinds of linters is that the former is designed to run on every {\emph{command}} and has access to the syntax tree of each command, while the latter is designed to run once on the ``final'' environment and only inspects the declarations present in the \href{https://leanprover-community.github.io/mathlib4_docs/Lean/Environment.html#Lean.Environment}{Environment}.
We refer to the former linters as ``syntax linters'' and to the latter linters as ``environment linters''. Both are semantic linters, in the sense that they can access the fully elaborated information and not merely the typed syntax.
Since both frameworks allow running arbitrary code, the distinction between them is mostly on what they are {\emph{expected}} to do and what they can do {\emph{easily}}, not on what they {\emph{could}} do.
Environment linters are good for {\emph{global}} validation (e.g.\ checking lemmas are in \href{https://lean-lang.org/doc/reference/latest/The-Simplifier/Simp-Normal-Forms/#simp-normal-forms}{\lean{simp} normal form}), but only warn you after the fact.
% XXX: \lean prints the dollar signs verbatim; we want to print them as formulas. \texttt is a small hack, but achieves this.
Syntax linters warn you right away and are best suited for {\emph{local}} validation (e.g.\ the valid notation \texttt{$\lambda$ x => x} is deprecated in favour of \texttt{fun x $\mapsto$ x}).

At the time of writing, there are 26 syntax linters defined in \mathlib and 10 more coming from its dependencies, as well as 17 environment linters, all but two defined in \verb|Batteries|.
We mostly focus on the \emph{syntax} linters: virtually all the environment linters have been described previously \cite{VanDoornEbnerLewis20}.%
\footnote{With the advent of Lean~4, the need for the typeclass linters described there has mostly disappeared: typeclass inference was substantially redesigned in Lean~4~\cite{SelsamUllrichDeMoura20}, taking into account the insights from \mathlib development. This means some linter warnings have become compiler errors, and the others have become less relevant and did not get reimplemented in Lean~4.}
To summarize more coherently how \mathlib uses linters, let us group them into general categories and highlight a few examples from each category.

\subsubsection{Deprecation linters} As mentioned in the previous section, linters help dealing with deprecations. Breaking changes could arise from certain tactics becoming more powerful or more efficient, making other tactics obsolete or undesirable. In such situations, the replaced tactics would be deprecated.
The \lean{deprecatedSyntax} linter warns about the use of deprecated syntax (e.g.\ a tactic), and informs the user what the new preferred expression is.

% This is a situation where the {\emph{syntax}} linters, as opposed to the {\emph{environment}} linters really shine: simply inspecting the environment, it is virtually impossible to recover which tactic produced which proof term. However, the {\emph{syntax}} of the declaration clearly contains this information very prominently.

\subsubsection{Style linters}
\Mathlib has a fairly detailed style guide (which helps to settle discussions about individual formatting preferences efficiently).
\Mathlib's style linters enforce various aspects of coding style and formatting, such as line length, whitespace conventions or the formatting of documentation comments.
This allows reviewers to focus their time on higher-level aspects, and helps contributors learn about \mathlib style without having to read all style rules at once.
% All of these are implemented as syntax linters.
We also enforce that every \mathlib file (that is not an \lean{import}-only file) contains a valid copyright header and a module doc-string summarising the content of the file.

\subsubsection{Import structure linters}
Keeping track of \mathlib's import structure is important for organisation and compilation speed.
A Lean module can only be compiled once all its dependencies have been fully checked\footnote{Lean's upcoming module system will change this trade-off, if and when it is adopted in mathlib. (Doing so is desirable in principle, but requires a lot of conscious library design work.)}: thus, a reasonable import structure enables parallelism and faster overall compilation. It also speeds up local development, as changing multiple files requires less recompilation, and makes it easier to place a new mathematical result.
\Mathlib has two main tools to help with keeping the import structure organised:
% MR: edited to omit the \lean{header} linter
Mario Carneiro's \texttt{shake} tool warns about unused imports, and the \lean{directoryDependency} linter warns about unintentional dependencies between different areas of \mathlib.

\subsubsection{Code pruning}
Another class of linters highlights redundant code.
For instance, the \lean{unusedVariable} linter warns when a variable in scope (such as a hypothesis in a proof) is not used.
While using Lean as a proof assistant, this helps catch ``free'' generalizations of theorems,
where a hypothesis that seemed to be needed turns out to be superfluous.
While using Lean as a general purpose programming language, this is more often useful to catch bugs,
as an unused variable may arise from an incorrect reference, or the omission of part of a case analysis.

Linters with similar scope are the \lean{unusedSeqFocus}, the \lean{unusedTactic}, and the \lean{unreachableTactic} linters,
each highlighting different ways in which a tactic proof can be simplified or streamlined.

\hide{There are also linters that make sure that if a \lean{set_option} command is issued,
that it really is necessary, in the sense that omitting it would cause some warning or error.
A \lean{set_option} command may become redundant with an improvement of the core code:
for instance, a proof may become substantially shorter, thanks to better algorithms being developed and hence increasing the default time-out may become unnecessary.

There is also a prototype of a linter that checks which environment \lean{variable}s have been declared, but never used.
Introducing ``unnecessary'' \lean{variable}s is fairly common when moving code between files:
some of the ``old'' \lean{variable}s may only have been used in the code that is being moved, but it may be hard to notice this, without a linter warning.}

\subsubsection{Tracking issues}

Certain commands are expected to be rarely used, or are only temporary adaptations.
For instance, increasing the time limit for type-checking is something that may be necessary for an individual declaration,
but increasing the limit for a whole file is likely a symptom of some deeper issue that should be addressed.
Hence, a command such as \lean{set_option maxHeartbeats 5000000} should be limited in scope to a single declaration.
The \lean{setOption} linter does exactly that: it flags uses of \lean{set_option maxHeartbeats x} suggesting to use \lean{set_option maxHeartbeats x in ...} instead.

Similar linters exist for \lean{open Classical}, which can hide issues that will make the code less usable,
and the \lean{globalAttributeIn} linter that warns that the scope of certain attributes is not ``local'', even in the presence of the (usually) scope-limiting \lean{in} keyword.

\subsection{Maintenance and code readability}

The linters in this category are specifically geared to improve maintainability of code and readability.
Since \mathlib evolves very quickly, code breakage is fairly common; any tooling which can diagnose and prevent such breakages is immensely helpful.
The linters in this section try to be pro-active about code breakage by reducing code brittleness.

\paragraph{Rigid and flexible tactics}
The \lean{flexible} linter starts from the premise that tactics exhibit two general behaviours:
``flexible'' tactics (such as \lean{simp}, \lean{norm_num} or \lean{aesop}) apply to a wide variety of goals and return similarly varied processed goals;
``rigid'' tactics (such as \lean{rw} or \lean{apply}) perform some very specific operations and may stop working even with very small changes of the current goal state.
% \begin{itemize}
% 	\item there are the ``flexible'' tactics (such as \lean{simp}, \lean{norm_num}, \lean{ring} or \lean{omega}) that apply to a wide variety of goals and return similarly varied processed goals;
% 	\item there are the ``rigid'' tactics (such as \lean{rw} or \lean{apply}) that perform some very specific operations and may stop working even with very small changes of the current goal state.
% \end{itemize}
% The prototypical example of a ``flexible'' tactic is \lean{simp}. Other flexible tactics are \lean{norm_num}, \lean{ring}, \lean{omega}, \lean{aesop}, and so on.
% The prototypical example of a ``rigid'' tactic is \lean{rw}. Other right tactics are \lean{exact}, \lean{refine}, \lean{apply}.

The \lean{flexible} linter tracks hypotheses and goals in each proof, following them around as various tactics are successively applied.
Whenever a ``flexible'' tactic modifies a hypothesis or a goal, no ``rigid'' tactic is supposed to modify that same hypothesis or goal:
indeed, the behaviour of the flexible tactic may change, e.g.\ because a new \lean{simp} lemma allows further simplification, making the following \lean{rw} fail.
Thus, preventing rigid tactics from following flexible ones helps with maintainability.
This question has some nuance: for instance, some flexible tactics (such as \lean{ring} or \lean{omega}) are ``normalizing'': even though they take a variety of inputs, their output is more predictable. Therefore, rigid tactics following a normalizing flexible tactic are allowed.

\paragraph{Handling multiple side-goals} In terms of goal management and readability, the \lean{multiGoal} linter warns as soon as
a tactic produces more than one side-goal, unless the next tactic is the focusing dot $\cdot$. % \lean{·}.
The code style enforced by this linter makes both the number of side-goals that remain after each tactic application as well as which parts of the code deal with which part of the proof completely transparent.
This helps with maintenance, since it limits code breakage to easily identified branches, rather than a generic total failure.
At the same time, it also improves code readability and structure, since the various arms of the proofs are visually differentiated.

\hide{The \lean{minImports} linter is another tool that helps code maintenance.
This linter inspects each declaration and, whenever a new declaration depends on more imported modules than the declaration preceding it, the linter emits a warning with a quantitative measure of the import increase.

This helps identify how to split files, based on their dependencies, providing a natural coalescence metric in terms of common prerequisites.
A similar tool is provided by the \lean{upstreamableDecl} linter, acting on a more granular level, giving information about individual declarations all of whose prerequisites are in an earlier file.}

\paragraph{Closing sections} Finally, let us mention the \lean{missingEnd} linter: it warns on \lean{section} commands that do not have a matching \lean{end} closure.
This addresses a maintainability issue: when moving code from one file to another, it is easy to miss the closing of a \lean{section}, producing a dangling one.
As a consequence, the scope of the currently non-\lean{end}ed section is persisted where it was not intended to be---and further hypotheses or declarations might be in scope. This can cause issues that are hard to diagnose, such as changing the meaning of statements or breaking proofs.

\subsubsection{Helpful and pedagogical suggestions}

Some linters attempt to preempt possible misunderstanding or common pitfalls.
The \lean{haveLet} linter flags uses of \lean{have} when introducing a non-\lean{Prop} assumption and
similarly uses of \lean{let} when introducing an assumption in \lean{Prop}.
Especially using \lean{have} for \lean{let} can cause confusion when the actual term introduced by the tactic is important for the proof.
% Note that the linter is only active on incomplete proofs in order to avoid continuously emitting warnings.
% Indeed, if a declaration is complete, there is no need to modify its proof.

Another example is the prototype \lean{papercut} linter.
One common mistake is accidentally using subtraction of natural numbers (which truncates at \lean{0}).
The linter flags such subtraction when a positivity assumption is not in scope.

All linters defined in \mathlib are available to projects using \mathlib: any improvements thus benefit the broader ecosystem.
Projects can easily adopt all of \mathlib's linters (or any subset thereof they choose). This helps maintaining code standards and styling conventions, and makes the ecosystem more homogeneous.

\hide{\subsubsection{Prototypes}
\begin{itemize}
\item MetaTest
\item Relative
\end{itemize}}

\section{Speeding up \mathlib}
\label{sec:speed}

% \Mathlib currently takes around 45 minutes to build completely on the continuous integration machines generously provided by the Hoskinson Center for Formal Mathematics
% (and 20 minutes when compiled on 32 core speedcenter machines).
As \mathlib grows to encompass more declarations, the time to build it will also increase.
% Some comments on how lines of code correlate to build times?
Keeping build times manageable requires constant attention to sources of slowdowns and inefficiencies.
The \mathlib \href{https://speed.lean-lang.org/mathlib4/home}{Speedcenter} % Or is there a better way to cite this?
is an instance of the benchmarking server \href{https://github.com/IPDSnelting/velcom}{VelCom}
that builds and benchmarks all commits to the master branch of \mathlib.
The benchmarks include overall measurements for Lean components such as the simplifier or typeclass inference,
and per-file compilation timings.
In addition to wall-clock (overall time from start to finish) and task-clock (takes into account parallelism) times measured in seconds,
most benchmark results are measured in instructions,
since the number of instructions is more stable on the benchmarking servers than the wall-clock time which is affected by process scheduling.
Continuous monitoring of build times is crucial to ensure no unexpected slowdowns occur.
Therefore, a bot reports notable outliers in benchmarks (changes of a measurement by 5\% or more) on the Lean community Zulip chat.
% Example?

\subsection{Speed and typeclasses} \label{subsec:speed-classes}
\Mathlib relies on the Lean typeclass system to perform automatic inference of many implicit arguments.
Typeclasses are flexible and have low infrastructure requirements on user code, but the flexibility also means designs have to be carefully consider to optimize their tradeoffs.
As the typeclass hierarchy grows, synthesis of instances becomes slower since a larger search space needs to be explored.
It is not only important to ensure synthesis succeeds quickly:
if synthesis fails, tactics can decide to fall back on an alternative proof strategy.
Lean's instance synthesis algorithm fails only after exploring the full search space,
and it is therefore especially important that the space does not grow too large.

One fundamental consideration in \mathlib's typeclass design is the level of \emph{bundling}~\cites{BaanenUseAndAbuse, Mathlib},
which concerns the way conjunctions of assumptions are modeled in classes.
For example, to represent ordered commutative monoids,
we can define a bundled class \lean{OrderedCommMonoid} that combines \lean{CommMonoid}, \lean{PartialOrder} and hypotheses stating the monoid operations respect the order structure.
Instead, unbundling the ordered algebra classes results in mixins combining the (existing) algebra class \lean{CommMonoid} with the (existing) order class \lean{PartialOrder} and adding in an extra \lean{IsOrderedMonoid} class.
Unbundling makes parameters more verbose: \lean{[OrderedCommMonoid M]} becomes \lean{[CommMonoid M] [LinearOrder M] [IsOrderedMonoid M]}.
As described with more detail in \cite{BaanenUseAndAbuse}, bundled design is preferred for its conciseness in \mathlib when possible.

% Unbundling algebra + order.
The speed tradeoffs changed when upgrading from Lean~3 to Lean~4.
A consequence of the bundled approach to typeclass design
is that the number of classes becomes multiplicative when extending along two distinct axes.
Taking the \lean{OrderedCommMonoid} class as an example formerly found in \mathlib,
this could then be extended in the order hierarchy to give \lean{LinearOrderedCommMonoid}, or in the algebraic hierarchy to give \lean{OrderedCommGroup}, or in both directions to give \lean{LinearOrderedCommGroup}.
When looking for a \lean{CommMonoid} instance, Lean would search through both the algebraic hierarchy and the ordered algebraic hierarchy.

Unbundling curbs this multiplicative growth by allowing expansion along individual axes.
This means to strengthen from ordered monoids to ordered groups, no additional ordering axioms are required,
and it suffices to replace a \lean{[CommMonoid M]} parameter with \lean{[CommGroup M]}.
Similarly, for linear ordered monoids we can replace \lean{[PartialOrder M]} with \lean{[LinearOrder M]}.
The result of unbundling is that the search through the one large search space is now three searches through smaller spaces,
and failing fast on one of these cancels the remaining searches too.
This \href{https://github.com/leanprover-community/mathlib4/pull/20676}{refactor by Yuyang Zhao} resulted in a 20\% decrease in typeclass inference time,
resulting in a 6\% overall speedup to \mathlib compilation.

\subsection{Speed and coercions}
The coercion system is another commonly used component of Lean.
Essentially, when a term \lean{t} is inferred to have type \lean{A}, where type \lean{B} is expected,
Lean attempts to find a suitable coercion, so the term can be replaced with a coerced version written $\uparrow$\lean{t}. % \lean{↑t}.
The coercion is found by synthesizing an instance of the typeclass \lean{CoeT A t B}.
Since coercions are triggered by type mismatches, this is a particular case of instance search that should fail fast.

Instances for \lean{CoeT} should not be declared directly, but rather chained together
from the \lean{CoeHead}, \lean{CoeOut}, \lean{Coe} and \lean{CoeTail} typeclasses.
The rules for which typeclass to instantiate for a particular coercion are subtle and not universally understood;
they are notably more complex compared to Lean~3.
As a consequence, a seemingly acceptable instance can cause a large performance regression.
For example, \lean{Coe A B} allows \lean{B} to be a free variable,
but doing so causes large performance regressions since such an instance always applies.%
\footnote{\url{https://github.com/leanprover-community/mathlib4/pull/22950}}

In a similar vein to Subsection~\ref{subsec:speed-classes}, large hierarchies extending coercion typeclasses should be avoided.
\Mathlib uses the \lean{FunLike} hierarchy to generalize over bundled homomorphisms. % Cite typeclasses paper?
Morphism classes such as \lean{MonoidHomClass} used to extend the \lean{FunLike} class,
which meant a failing coercion would first trigger a search through the entire \lean{FunLike} hierarchy.
Even worse, since \lean{MonoidHomClass} depends on \lean{Monoid} instances,
this also triggers a search through the algebraic hierarchy.
The solution was to unbundle inheritance from \lean{FunLike}, keeping only \lean{FunLike} and \lean{EquivLike}
and passing these classes as an instance parameter to every morphism class.
The \href{https://github.com/leanprover-community/mathlib4/pull/8386}{refactor by Anne Baanen performing this unbundling} resulted in a 33\% speedup for typeclass instance synthesis
and an overall 19\% decrease in build instructions.

\subsection{Speed and definitional equality}
Type checking, and definitional equality in particular,
is commonly invoked in the Lean elaboration and kernel checking process
and can result in arbitrarily complex computations:
terms can be large and their size can increase further when unfolding definitions referenced by the term.
Equality checks are intended to be ``cheap'' basic steps in implementing more complex routines such as typeclass instance synthesis,
so slowdowns in checking have a large impact.

Generally, definitions are designed to not require unfolding unless the unfolding is requested.
Each definition in Lean has a \emph{transparency level} controlling the circumstances in which it is unfolded.
Most tactics are designed to only unfold definitions explicitly marked as reducible,
with some tactics also unfolding at the default semireducible transparency level.
Marking definitions as irreducible prevents unfolding altogether,
skipping a potentially expensive equality check if it does not obviously succeed.

A particular example of complex definitional equality checking can be found in structure declarations.
Large structures can already be quite complex terms in themselves,
which can be hidden behind further definitions and constructions,
and the possibility of diamond inheritance can then result in nonobvious equality checks.
This all results in high levels of unfolding being required to verify the equality of terms.

Typeclasses in Lean are implemented as structures, so the complex typeclass hierarchy also has an impact here.
The \lean{Field} typeclass has 19 data-carrying structure fields that must be checked for equality.
By ensuring typeclass instances are already in a normal form, less unfolding will be needed.
Matthew Robert Ballard, Kyle Miller and Eric Wieser implemented the \lean{fast_instance} term elaboration macro
that normalizes instance declarations by unfolding definitions and replacing subterms with synthesized instances.
An initial implementation of this macro resulted in a 5\% decrease in type checking time
and halved compilation times for many files.%
\footnote{\url{https://github.com/leanprover-community/mathlib4/pull/20993}}

\subsection{Speed and tactics}

During the compilation of \mathlib, Lean spends approximately 14\% of its time running the \lean{simp} tactic,
and 7\% of the time invoking other tactics.
Tactics are diverse in the ways they are implemented and invoked, but we can attempt to find some rules of thumb for optimization.

An important change in Lean~4 was the addition of \emph{discrimination trees}~\cite{McCuneDiscrTrees, HandbookTermIndexing} that allow for efficient association of terms with values.
A discrimination tree is a tree datastructure whose branches are indexed by symbols (either a constant, or the special symbol \lean{*} representing free variables).
Values are stored at leaf nodes.
A closed term where each symbol has a fixed arity can be represented as a string of symbols;
this maps directly to a particular path in the tree.
Queries into a discrimination tree overapproximate the result, since all variables map to the same symbol \lean{*}.
Still, the approximation is much finer than previously in Lean~3, where terms were indexed by only the first constant that appeared~\cite{Lean4}.

The simplifier tries to repeatedly rewrite using a set of lemmas and tactics (called a \lean{simp}~set) on all subterms in an expression.
By querying a discrimination tree, it can quickly determine potentially relevant rewrites.
Optimization practices common in the Lean~3 version of \mathlib have become much less necessary after the port to Lean~4,
such as restricting \lean{simp} from the default large \lean{simp}~set to a smaller specific list
(using the \lean{simp only} keywords)
since many fewer lemmas have to be tested for applicability.
Going from \lean{simp only} to \lean{simp} lowers the maintenance burden,
since a change to the default \lean{simp}~set does not need to be copied to each involved \lean{simp only} call.

The \lean{simp}~sets still need to be restricted for efficiency if multiple lemmas could apply depending on different hypotheses.
For example, \lean{div_self a : a / a = 1} applies if \lean{a} is in a \lean{Group},
but a field does not form a multiplicative group.
Therefore, the \lean{field_simp} tactic specialized to simplifying expressions in the theory of fields
excludes such lemmas from the set for efficiency.%
\footnote{\url{https://github.com/leanprover-community/mathlib4/pull/21326}}

Another avenue to optimizing the use of tactics is to use weaker, faster tactics instead of more powerful but slower tactics.
The Aesop tactic provides general proof search~\cite{LimpergAesop}
and is also invoked as part of other, more specialized tactics to discharge goals.
The general nature of this proof search makes Aesop very powerful,
but this power is often unnecessary when the proof it needs to find has a predetermined format.
We have found that replacing powerful Aesop-based tactics like \lean{measurability}
with tactics that have a simpler strategy like \lean{fun_prop} is almost always possible and reduces running time appreciably.
Similarly, small tweaks such as applying local hypotheses or trying reflexivity to discharge an equality goal before starting Aesop
can lead to appreciable speedups.%
\footnote{\url{https://github.com/leanprover-community/mathlib4/pull/21330}}

\section{Tracking and addressing technical debt}
\label{sec:techdebt}

Ward Cunningham introduced~\cite{CunninghamTechDebt} the term \emph{technical debt} in 1992, writing:
\begin{quote}
	Shipping first time code is like going into debt.
	A little debt speeds development so long as it is paid back promptly with a rewrite.
	Objects make the cost of this transaction tolerable.
	The danger occurs when the debt is not repaid.
	Every minute spent on not-quite-right code counts as interest on that debt.
	Entire engineering organizations can be brought to a stand-still under the debt load of an unconsolidated implementation
\end{quote}
This concept is all the more important in a large foundational library like \mathlib,
especially in light of its cohesive and tightly integrated nature.
In this section we describe various types of technical debt we have encountered in \mathlib,
and how the community tracks and addresses them.

\paragraph{Porting notes}
From 2021--2023, the community ported \mathlib from Lean~3 to Lean~4.
This process ought to be documented in more detail elsewhere,
but for the purposes of the present discussion we summarize the points of interest.

For all practical purposes, Lean~4 is a distinct language from Lean~3---there is no backwards compatibility.
Mario Carneiro wrote a tool \texttt{mathport} that automatically translates Lean~3 code to Lean~4 code on a best-effort basis,
and a small group of people implemented the common tactics in Lean~4.
In October/November 2022,
this preparatory work transitioned into a wide-scale and coordinated effort to port the theory files of \mathlib.
Over the course of 10 months, the community ported on the order of a million lines of code, or roughly a dozen files per day.
% All the while, \mathlib~3 still accepted new contributions to unported files, while \mathlib~4 started accepting contributions to the ported files.

Naturally, there were many places where a faithful translation was not possible.
For the sake of speed, the porting team decided to liberally annotate issues with ``porting notes'' that described the problem and the workaround.
As a result, by the summer of 2023, the porting process was complete,
but \mathlib~4 contained well over 5000 porting notes.

Pietro Monticone made a valiant effort to organize and classify the porting notes in February and March of 2024.
Subsequently, Anne Baanen, Kim Morrison, and Johan Commelin processed and addressed over two thirds of the porting notes in the first quarter of 2025.
Many of these porting notes became obsolete because of new and expanded functionality in Lean~4.
At the time of writing, roughly 1500 porting notes remain.

\paragraph{Adaptation notes and backwards compatibility flags}
New versions of Lean~4 are released on a monthly basis.
\Mathlib maintains a branch that tracks the nightly releases, which simplifies migrating to the new version of Lean~4 when it is released.
This tracking branch provides feedback for the development of Lean~4:
if the daily transition to a new nightly version of Lean~4
requires some unsatisfactory workarounds,
then these are annotated with \texttt{\#adaptation\_note} comments.
In essence, these notes have the same functionality as porting notes.
Occasionnally, there is no simple solution for a problem that arises from a new version of Lean~4.
In such an event the old feature can be re-enabled via a backwards compatibility flag.
This allows the \mathlib community to transition to the new version of Lean~4 at a reasonable pace.

% Both adaptation notes and backwards compatibility flags are forms of technical debt.
% Their presence is tracked in a weekly report that is automatically posted to the Zulip chat of the Lean community.

\paragraph{Reporting technical debt}
The shell script \texttt{scripts/technical-debt-metrics.sh} is used to create a report of the current state of known technical debt in \mathlib.
This includes tracking adaptation notes, porting notes and backwards compatibility flags.
Continuous integration runs this script on a weekly basis, and posts the report (including the delta with respect to the previous week) to the Zulip chat of the Lean community.
Additionally, this report is computed for each pull request, and posted to a dedicated ``PR summary'' comment at the top of the pull request.
This allows the reviewer to quickly see the current state of technical debt in the pull request, and take this into account when deciding whether to approve the pull request.

\Mathlib also has a mechanism in place to surface pull requests addressing technical debt: this includes both a dedicated label and triage dashboard (see Section~\ref{sec:triageTooling}) and a Zulip stream where such PRs can be brought to attention.

One downside of tracking and reporting technical debt in this manner is indicated by Goodhart's Law:
``When a measure becomes a target, it ceases to be a good measure.''
In this context, this means that the community is inclined to focus on the technical debt that is being tracked,
whereas there may be other forms of technical debt that fly under the radar.
To alleviate this, there is a dedicated thread on the Lean community's Zulip chat listing less measurable forms of technical and organisational debt.
(One example of the latter is documenting unwritten informal \mathlib policy: sharing such knowledge enables other contributors to learn this for themselves and lowers the barrier towards contribution. An example of the former is reorganizing top-level directories: due to \mathlib's age, some of these are not coherent any more.)
Everybody can add items to the list, facilitating a collective understanding of future maintenance tasks.

\section{Scaling code review}
\label{sec:reviewTooling}

Since its beginning, the \mathlib library required code review prior to merging changes.
This is enforced through the \href{https://bors.tech/}{bors} bot: pull requests are merged automatically upon approval by an authorised user (i.e., a \mathlib maintainer).%
\protect\footnote{Using bors also enforces the not rocket science principle~\cite{GraydonNotRocketScience}: ensuring that the main (development) branch of \mathlib always passes all continuous integration tests.}

This policy has several reasons. For one, formalisation in Lean ensures a strong level of correctness: a formalised mathematical proof which compiles without any \lean{sorry}\footnote{and is not intentionally manipulating the Lean environment or adding additional axioms---both of which are checked for by \mathlib's continuous integration}
will be a correct and complete proof. However, designing an integrated mathematical library \emph{well} still requires expert knowledge: usually, there are different potential designs, with vastly different usability and ergonomics. Code review is one way to form and share common knowledge about design best practices.
It also has an empowering function: new contributors can propose new code, without being afraid of breaking other proofs (if their code compiles, the proofs are still fine) or overlooking important aspects---because an experienced Lean user will look over their contribution. This way, code review is a way of teaching and onboarding new contributors.

\Mathlib is growing steadily, as is its influx of new contributions. At the same time, many of its contributors are volunteers or work on \mathlib as a side project. Most of \mathlib's maintainers have teaching, administrative or supervision commitments: allowing \mathlib to scale requires lightening their work.
This has both social and technical aspects.
On the social side, maintaining \mathlib means embracing teamwork and delegating responsibility. Approval of \mathlib pull requests is a multi-stage system, allowing many people to contribute.

Code review in \mathlib is decentralised: everybody is allowed and encouraged to review contributions; there are \href{https://leanprover-community.github.io/contribute/pr-review.html}{guidelines} teaching what to review for. This ranges from more basic aspects (such as adherence to \href{https://leanprover-community.github.io/contribute/style.html}{\mathlib's style guidelines}, or whether new documentation is comprehensible) to advanced aspects like the correct code location and library design patterns.
There are regular review meetings, where reviewers and interested contributors review code together---both to make this work more enjoyable, and to mentor potential new reviewers.

\Mathlib's maintainers have the final say on approving a pull request. The number of maintainers keeps growing steadily, and is currently around 30. Maintainers have different specialisations; together, they cover the breadth of \mathlib. Having multiple knowledgeable maintainers in one area is beneficial, so they can discuss difficult decisions.

Between maintainers and all contributors is the group of \emph{\mathlib reviewers}. These are \mathlib contributors who have provided useful reviews in the past. A reviewer can approve a pull request by commenting \texttt{maintainer merge}, which notifies the maintainers. Usually, these get merged (or at least discussed further) more quickly. This system is a bit like journal referees and editors: the reviewer group lightens the load on maintainers; the direct notification mechanism provides an additional small incentive for reviewers.
This works well in practice: in the last year, about one in three merged pull requests was approved by a reviewer before they were merged.

On a technical level, we can save reviewer and maintainer time by automating tedious or mechanical aspects of this process. Two basic means of automation are using continuous integration with \emph{bors} and automatic maintainer notifications through the \texttt{maintainer merge} command.
Let us discuss \mathlib's review automation from two further angles: tooling for reviewers and tooling for triage.

\subsection{Tooling for reviewers}

There are multiple ways for automation to support pull request reviewers.
To begin with, we need to find a suitable reviewer for a given pull request. A pull request with a new theorem is best reviewed by somebody familiar with the theorem's area; something similar holds for a contribution improving \mathlib's infrastructure, adding a new linter or adding a new tactic.
Each new pull request adding new content is automatically labelled according to its area. There is a file collecting each reviewer's area of interest and expertise. This allows automatically suggesting suitable reviewers for every pull request---taking their expertise and reviewing capacity into account. At the time of writing, this automatic assignment is run on an experimental basis.
This labelling information has also been used to identify missing reviewers: if pull requests in a certain area take a long time to get reviewed, recruiting further reviewers could be worthwhile.

Secondly, we have created custom tooling exposing important information to reviewers. When a pull request is filed, a script creates a sticky comment summarising key information about a PR. This comment is updated in-place as the pull request is amended. The comment shows which lemmas were renamed (so reviewers can spot those more easily, and check if deprecation warnings are present). It also exposes when transitive imports of files change, which is extremely helpful for reviewing import refactoring. Finally, the summary comment shows any changes in technical debt metrics.

Further, there is substantial automation to help with information flow within the project.
Collaboration on \mathlib occurs primarily on two sites: GitHub, where git repositories including \mathlib and its dependencies are hosted, and the leanprover-community \href{https://leanprover.zulipchat.com}{Zulip}, a moderated chat site where community members discuss their contributions and other Lean-related topics.
Automation can help with bridging these two venues and keeping some information synchronised.
Whenever a specific pull request is mentioned in the Zulip chat, a script posts an emoji reaction indicating the current pull request status (such as, whether it is waiting for author changes, a review or was already merged). This allows reviewers and maintainers to see at a glance if action has already been taken. This reaction is updated by a GitHub actions script whenever the pull request's status changes.

We make extensive use of \href{https://zulip.com/help/add-a-custom-linkifier}{Zulip's custom linkifier feature} to allow users to link directly to issues or pull requests in the \mathlib repository, \mathlib documentation entries, and other useful pages.

% omitted for brevity
% At the scale of \mathlib, the volume of GitHub notifications is quite high (up to dozens of alerts per hour, at peak hours), so having select events announced on Zulip ensures that important maintenance work is not missed. For example, the status of pull requests updating \mathlib to nightly versions of Lean as well as automated updates of dependencies are posted automatically on Zulip, alerting the community more quickly to failures.
% As mentioned in Section~\ref{sec:techdebt}, technical debt reports are also compiled and posted to Zulip weekly.

\subsection{Triage tooling}
\label{sec:triageTooling}

The number of open \mathlib pull requests is growing steadily. At the time of writing, there are about 1500 open pull requests.
Often, new pull requests are getting reviewed quickly, but their sheer volume makes it difficult to keep track of every single PR.
Better tooling helps to ensure that every pull request is kept track of and receives a timely response.

Investigating this closely shows that different users of such tooling have differing needs. For instance, a reviewer may want to find reviewable pull requests matching their areas of interest. \Mathlib's maintainers would like to know about pull requests which were already approved by a reviewer.
Other contributors may be interested in recent pull requests to ``see what is going on'', or viewing all pull requests related to a particular area. Pull request authors may want to see the status of all their pull requests or find out why their pull request is not on the review queue.
Most importantly, a triage team (see below) keeping track of pull requests' state needs better tooling to see the state of each pull request.

GitHub's default interface can do many of the above tasks; however, at \mathlib's scale we have increasingly run into its limitations. For example, to triage reviewable pull requests one would like to know the total time a pull request is already waiting for review. This need not be a continuous time span, so computing this needs to consider a pull request's change over time. GitHub only exposes a ``last updated'' date, which does not provide this. (In addition, this ``last update'' measures almost every activity, which is often too sensitive for our purposes: a bystander's comment or a notification of a merge conflict resets the last activity counter, but need not mean review activity.) Similarly, sorting pull requests by their total diff can be useful information, as is the number of PRs depending on a given PR.

\begin{figure}
\includegraphics[width=\textwidth]{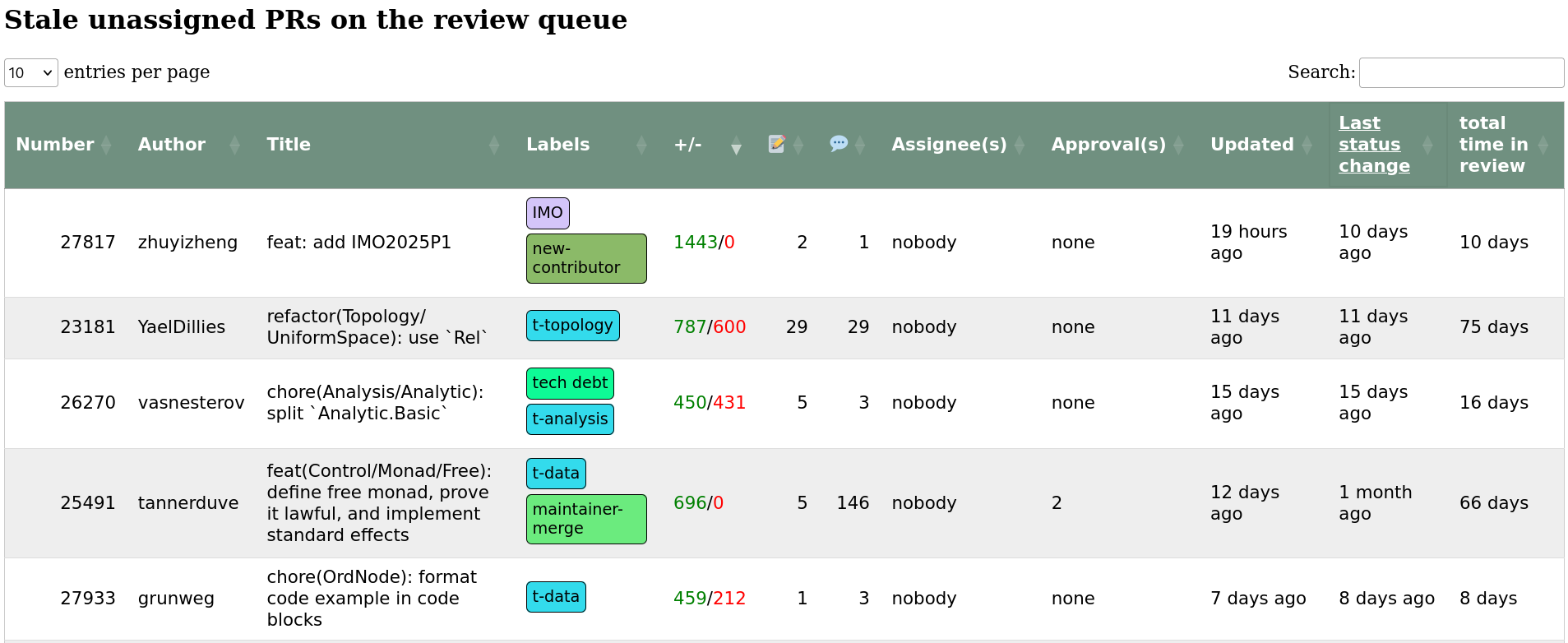}
\caption{A typical dashboard from \mathlib's review and triage dashboard}
\label{fig:queueboard}
\end{figure}

We have built a custom editorial tool to address all these shortcomings.
Pull requests are grouped by their status (such as waiting for review, blocked on another pull request or waiting for the author to respond to review comments). These categories are tailored to \mathlib's infrastructure (for instance, pull request failures related to \mathlib's infrastructure are highlighted automatically, speeding up the diagnosis and fix).
In each category, we compactly display (see Figure~\ref{fig:queueboard}) all pull requests together with additional meta-information such as its author, title and labels, but also additional information such as diff size or the total time in review. These lists can be semantically sorted by each category.
This tooling can adapt and grow with \mathlib's needs easily.

We ran a survey among \mathlib contributors, reviewers and maintainers. The results provided many useful suggestions for improving this tool further (many of which have already been implemented).
The survey result also affirms that this changes how contributors look for pull requests: the dashboard has become part of many maintainers' standard workflow.
The triage functionality in particular is used by maintainers, but also by the newly-founded triage team: having a common webpage to look at makes this process a lot smoother.

Under the hood, this dashboard has two components.
One keeps a cache, containing the metadata we care about for every open pull request. Periodically, we query GitHub's \verb|graphql| API to update this data for any pull requests which received updates, in effect creating a database of each pull request's relevant state. The code for this amounts to a few hundred lines of shell and Python scripts: this could easily be run on a local computer, coupled with a cronjob.

The other component is responsible for the final user interface. It processes the downloaded metadata to extract the aggregate data we care about. (For instance, from the history of all label changes on a pull request, we determine the total time that PR was ready for review.) Using this metadata, we create the triage dashboard. The dashboard is a static webpage, using jQuery datatables to create the dashboards in each category: these can be sorted and filtered dynamically. Using a static webpage means fast response times (filtering a list of pull requests is almost instantaneous), but means the webpage needs to be manually refreshed to see updated changes.
Altogether, this project consists of about 4500 lines of Python code, and about 1000 lines of shell scripts, \texttt{.graphql} schemas etc.

\section{Conclusion}

Growing \mathlib would be impossible without automation to scale its maintainance: this applies on the social, organisational and technical levels.
While we cannot measure its effects precisely, the fact that \mathlib still grows fast shows how useful it is.
For instance, \mathlib's deprecation system has made updating to a newer version of \mathlib a much smoother process.
Currently, such updates still require manual work, but this task can be automated well: both for deprecated declarations and modules, there are prototype tools performing the replacement automatically.

An auto-formatter for Lean would improve dealing with code style questions significantly.
Lean's extensibility makes this difficult in general: users can define new syntax and custom notation, which can drastically change the look and feel of the language.\footnote{For example, one can allow users to write in a form of controlled natural language \cite{Massot24}: formatting that would be a very different endavour.}
Formatting user-defined syntax and notation requires knowing the preferred formatting for these.
While \mathlib makes extensive use of custom notation to make its code resemble mathematical notation, this is relatively benign: all syntax has an associated pretty-printer which indicates some reasonable formatting.

% The code for \mathlib's review and triage dashboard is open source.
\mathlib's review and triage dashboard is mostly \mathlib-agnostic: the particular PR categories are tuned to \mathlib, but the underlying logic is not.
With appropriate configuration, other open source projects could use this as well.

\subsubsection*{Acknowledgements}
We would like to thank the anonymous referees for their helpful comments and suggestions.
Michael Rothgang was funded by the Deutsche Forschungsgemeinschaft (DFG, German Research Foundation) under Germany's Excellence Strategy -- EXC-2047/1 -- 390685813.
This material is based upon work supported by the U.~S.\ National Science Foundation Award No.\ DMS-2302263.

\printbibliography
\end{document}